\documentclass[conference]{IEEEtran}
\IEEEoverridecommandlockouts
\usepackage{cite}
\usepackage{amsmath,amssymb,amsfonts}
\usepackage{algorithmic}
\usepackage{mathtools}
\usepackage{graphicx}
\usepackage{subcaption}
\usepackage{hyperref}
\usepackage[american]{circuitikz}
\usepackage{multirow}
\usepackage{textcomp}
\usepackage{xcolor}
\usepackage{tikz}
\usetikzlibrary{shapes,arrows}
\usepackage{verbatim}

\def\BibTeX{{\rm B\kern-.05em{\sc i\kern-.025em b}\kern-.08em
    T\kern-.1667em\lower.7ex\hbox{E}\kern-.125emX}}
    
\begin{document}

\tikzstyle{block} = [draw, rectangle, 
    minimum height=2em, minimum width=4em]
\tikzstyle{sum} = [draw, circle, node distance=1cm]
\tikzstyle{input} = [coordinate]
\tikzstyle{output} = [coordinate]
\tikzstyle{pinstyle} = [pin edge={to-,thin,black}]

\title{Switched Control Applied to a Totem-Pole Bridgeless Rectifier for Power Factor Correction\thanks{The authors thanks Fundação de Amparo à Pesquisa e Inovação do Estado de Santa Catarina (FAPESC) - Grant number 288/2021 for the financial support.}
}

\author{\IEEEauthorblockN{Luan Vinícius Fiorio}
\IEEEauthorblockA{\textit{Department of Electrical Engineering} \\
\textit{Santa Catarina State University}\\
Joinville, Brazil \\
luan.lvf@edu.udesc.br}
\and
\IEEEauthorblockN{Tiago~Jackson~May~Dezuo}
\IEEEauthorblockA{\textit{Department of Electrical Engineering} \\
\textit{Santa Catarina State University}\\
Joinville, Brazil \\
tiago.dezuo@udesc.br}
\and
\IEEEauthorblockN{Yales~Rômulo~de~Novaes}
\IEEEauthorblockA{\textit{Department of Electrical Engineering} \\
\textit{Santa Catarina State University}\\
Joinville, Brazil \\
yales.novaes@udesc.br}
}

\maketitle

\begin{abstract}
The wide range of operation of bridgeless rectifiers requires a control technique that guarantee robustness. Linear Power Factor Correction (PFC) control techniques, although effective, cannot guarantee such robustness. Nonlinear techniques such as one cycle control are more robust, but other options should be explored. In this work, an affine model is obtained for a Totem-Pole Bridgeless Rectifier (TPBR). An extension to an existing switched control design technique is presented in order to achieve PFC in a robust fashion for the TPBR. Simulations with nonideal components and distorted grid voltage show a precise, fast and robust performance of the switched controller. The effective reference following of the proposed method allows the user to define a current reference waveform that prioritize THD or power factor, depending on the application and norm requirements.

\end{abstract}


\section{Introduction}
Bridgeless rectifiers, as the Totem-Pole Bridgeless Rectifier (TPBR) \cite{Salmon1995}, can achieve higher efficiency than the conventional Boost converter and are able to reduce conduction losses. The limitation in those topologies were given by the semiconductor technology \cite{Huber2008}, mainly the reverse recovery of body diodes of the MOSFETs. With the increase of semiconductors research and products, high-performance silicon Insulated-Gate Bipolar Transistor IGBT \cite{Wu2017}, wide bandgap semiconductors like Silicon Carbide (SiC) \cite{Zhu2016,Wu2017} and Gallium Nitride (GaN) \cite{Wu2017,Zhang2018} have allowed the TPBR to achieve high efficiency in practice, up to 99 \%.

Possible control strategies that are able to control the TPBR with power factor correction for the input current are: Average-Analog (A-A) \cite{UC3854}, Digital-Average (D-A) \cite{Buso1998}, Trailing-Edge Modulated One Cycle Control (TEM-OCC) \cite{Lai1998,Brown2005}, Digital-Peak Current (D-PC) \cite{Wu2017}, Leading-Edge Modulated OCC (LEM-OCC) \cite{Lai1998}, LEM-OCC Stability and Distortion (LEM-OCC-SD) \cite{Fischer2020}, and LEM-OCC-SD Simplified (LEM-OCC-SDS) \cite{Fischer2020}. The OCC-based methods have shown to be more robust than the previously cited techniques, which is expected from its nonlinear nature.

Another nonlinear approach that could achieve high power factor is Switched Control \cite{Branicky1998,Colaneri2008}. As the TPBR can be modeled as an affine\footnote{Nonlinear systems that are linear in the input.} switched system, a switching rule design \cite{Trofino2009,Trofino2011} can be obtained in order to stabilize the process that guarantees Lyapunov-based stability for all designed operation range, which presents stronger theoretical robustness guarantees than OCC-based approaches for any uncertain parameter varying inside a polytope region \cite{Battistelli2017}. The switching control can also be extended to include optimization objectives, as the $\mathcal{H}_{\infty}$ method for minimizing the effects of disturbance \cite{Trofino2012}. As the switched control laws are designed in an offline fashion, its application can be done with a digital signal processor with low computational complexity.

This paper presents the application of a switching rule to a Totem-Pole Bridgeless Rectifier operating in Continuous Conduction Mode (CCM) with the objective of stabilization and power factor correction. Extensions are made to the design procedure of the switching rule to allow power factor correction. Section~\ref{sec:tpbr} presents the modeling of the TPBR as a switched system, as well as a brief review over the existing Power Factor Correction (PFC) control techniques. Section~\ref{sec:switched} reviews the switching rule design considering the switched control approach for affine systems as presented in \cite{Trofino2011}, whilst Section~\ref{sec:pfc} extends the existing design procedure to achieve a high power factor over rectifier type of converters. Finally, Section~\ref{sec:results} presents the obtained nonideal simulation results and Section~\ref{sec:conclusions} concludes this work.

\section{The Totem-Pole Bridgeless Rectifier}
\label{sec:tpbr}
\begin{figure}[!t]
    \centering
    \begin{circuitikz}[scale=0.75]
    \ctikzset{
        bipoles/length=.8cm
    }
        \draw
        (-0.5,1.6)   to[vsourcesin, v=$v_{in}$]      (-0.5,0)
        (-0.5,1.6)   to[inductor, i>=$i_{L_B}$, -*, label=$L_B$]     (2,1.6)
        (2,2)   node[nigbt, bodydiode, anchor=E](S1){$S_1$} (2,2)
        (2,0)   to[short]           (2,2)        
        (2,0)   node[nigbt, bodydiode, anchor=C](S2){$S_2$} (2,0)
        (S1.C)  to[short]           (2,4)
        (2,4)   to[short]           (4,4)
        (4,4)   to[short,*-] (4,3.5)   to[diode, l=$D_1$, invert]  (4,2)
        (4,2)   to[short]           (4,0)
        (4,0)   to[diode, l=$D_2$, invert]  (4,-1.5) to[short,-*] (4,-2)
        (S2.E)  to[short]           (2,-2) -- (4,-2)
        (-0.5,0)    to[short,-*]    (4,0)   
        (4,4)   to[short]           (6,4)
        (6,4)   to[capacitor, label=$C_o$, v = $v_{C_o}$, *-*]  (6,-2)
        (4,-2) -- (6,-2)
        (6,4) -- (8,4)
        (6,-2) -- (8,-2)
        (8,4)   to[resistor, label=$R_L$]   (8,-2)
        (8,-2) to[short,*-]         (8,-2)
        (8,-2) node[ground]{}       (8,-2)
        ;
    \end{circuitikz}
    \caption{Totem-pole Bridgeless Rectifier.}
    \label{fig:tpbr_c}    
\end{figure}
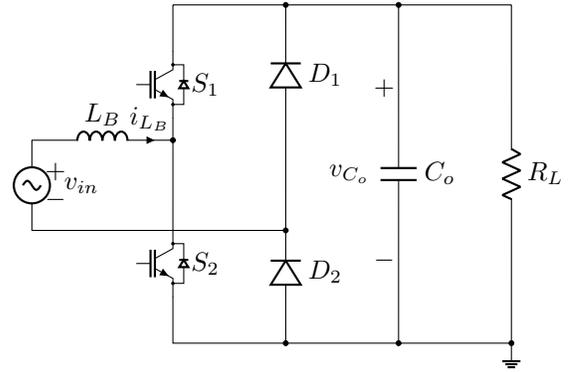

The Totem-Pole Bridgeless Rectifier (Fig.~\ref{fig:tpbr_c}) is an AC-DC converter with a totem-pole arm of transistors in the rectifier bridge. The converter is based on the full-bridge diode rectifier with capacitive filter but it uses switches in the place of two diodes, which has the potential to reduce conduction losses and allows the designer to achieve high power factor using a PFC control technique \cite{Huber2008}.

The modes of operation for the TPBR rely directly on the sign of the input voltage and are presented in Table~\ref{tab:modes}. At the first mode, the current flows through the $S_2$ and $D_2$, with the output voltage $v_{C_o}$ being held by the energy stored in capacitor $C_o$. With the command to block $S_2$, the second mode begins, but the current now flows through $D_2$ and the intrinsic diode of $S_1$ to the output capacitor and load. The third and fourth modes follows the same behavior, analogously. The described operation regards CCM operation, and it used as base for the affine model design in next section.

\begin{table}[!t]
\renewcommand{\arraystretch}{1.3}
\caption{Modes of operation of the TPBR.}
\label{tab:modes}
\centering
\begin{tabular}{c c c c}
\hline
\textbf{Mode} & \textbf{$\mathbf{S_1}$ state} & \textbf{$\mathbf{S_2}$ state} & \textbf{$\mathbf{v_{in}}$ sign}\\
\hline
1 & OFF & ON & $> 0$ \\
2 & OFF & OFF & $> 0$ \\
3 & ON & OFF & $< 0$ \\
4 & OFF & OFF & $<0$ \\
\hline
\end{tabular}
\end{table}

\subsection{Affine model}
\label{ssec:tpbr_model}
The system can be represented in a nonlinear fashion by considering the state-space affine model \cite{Trofino2011}
\begin{equation}
    \dot{x}(t) = A_i x(t) + b_i, \ i \in \mathcal{M} \coloneqq \{1,...,m\},
    \label{eq:affine}
\end{equation}
where $x \in \mathbb{R}^{n}$ are the states with its derivatives $\dot{x}(t)$, $A_i \in \mathbb{R}^{n \times n}$, $b_i \in \mathbb{R}$, are the matrices for each mode $i \in \mathcal{M}$ of the system.

Considering $x = [i_{L_B} \ v_{C_o}]'$ as the state vector of the TPBR, the expressions for its model in CCM can be obtained as follows: for Mode~1, a mash with the inductor $L_B$, the conducting switch $S_2$, the diode $D_2$ and the input voltage source $v_{in}$ results in the differential equation
\begin{equation}
    \frac{d i_{L_B}}{d t} = - i_{L_B} \frac{R_B}{L_B} + v_{in} \frac{1}{L_B},
\end{equation}
where $R_B$ is the inductor's resistance. Also, a node at the output capacitor $C_o$ and the load resistance $R_L$ results in
\begin{equation}
    \frac{d v_{C_o}}{d t} = - v_{C_o} \frac{1}{R_L C_o}.
\end{equation}
Nevertheless, resulting in the state-space matrices for the first mode:
\begin{equation}
    A_1
    = 
    \begin{bmatrix}
        -\frac{R_B}{L_B} & 0 \\
        0 & -\frac{1}{R_L C_o}
    \end{bmatrix}
    , \
    b_1 =
    \begin{bmatrix}
        \frac{v_{in}}{L_B} \\
        0
    \end{bmatrix}
    .
\end{equation}

In Mode~2, the current flows through diode $D_2$ and the intrinsic diode of switch $S_1$ to the output capacitor and load, which gives
\begin{equation}
    \frac{d i_{L_B}}{d t} = - i_{L_B} \frac{R_B}{L_B} - v_{C_o} \frac{1}{L_B} + v_{in} \frac{1}{L_B},
\end{equation}
and
\begin{equation}
    \frac{d v_{C_o}}{d t} = i_{L_B} \frac{1}{C_o} - v_{C_o} \frac{1}{R_L C_o},
\end{equation}
being represented as
\begin{equation}
    A_2
    = 
    \begin{bmatrix}
        -\frac{R_B}{L_B} & -\frac{1}{L_B} \\
        \frac{1}{C_o} & -\frac{1}{R_L C_o}
    \end{bmatrix}
    , \ 
    b_2
    =
    \begin{bmatrix}
        \frac{v_{in}}{L_B} \\
        0
    \end{bmatrix}
    .
\end{equation}

During Mode~3, the input current flows through the switch $S_1$ and the diode $D_1$. The output voltage $v_o$ is held steady by the output capacitor $C_o$, subject to its ripple. This operation can be translated to the expressions
\begin{equation}
    \frac{d i_{L_B}}{d t} = - i_{L_B} \frac{R_B}{L_B} + v_{in} \frac{1}{L_B},
\end{equation}
and
\begin{equation}
    \frac{d v_{C_o}}{d t} = - v_{C_o} \frac{1}{R_L C_o},
\end{equation}
resulting in the state matrices
\begin{equation}
    A_3
    = 
    \begin{bmatrix}
        -\frac{R_B}{L_B} & 0 \\
        0 & -\frac{1}{R_L C_o}
    \end{bmatrix}
    , \
    b_3 =
    \begin{bmatrix}
        \frac{v_{in}}{L_B} \\
        0
    \end{bmatrix}
    .
\end{equation}

Mode~4, similarly to Mode~2, occurs when both switches are in OFF-state and the current flows through the intrinsic diode of $S_2$ and diode $D_1$ to the output capacitor $C_o$ and load $R_L$. The expressions that describe this behavior are
\begin{equation}
    \frac{d i_{L_B}}{d t} = - i_{L_B} \frac{R_B}{L_B} + v_{C_o} \frac{1}{L_B} + v_{in} \frac{1}{L_B},
\end{equation}
and
\begin{equation}
    \frac{d v_{C_o}}{d t} = - i_{L_B} \frac{1}{C_o} - v_{C_o} \frac{1}{R_L C_o},
\end{equation}
resulting in
\begin{equation}
    A_4
    = 
    \begin{bmatrix}
        -\frac{R_B}{L_B} & \frac{1}{L_B} \\
        -\frac{1}{C_o} & -\frac{1}{R_L C_o}
    \end{bmatrix}
    , \ 
    b_4
    =
    \begin{bmatrix}
        \frac{v_{in}}{L_B} \\
        0
    \end{bmatrix}
    .   
\end{equation}

\subsection{Control strategies for power factor correction}

%

Average-Analog aims to control average values, which can result in satisfactory THD under light load operation, but the average nature makes the control slower and less precise. Digital-Average uses rising edge sampling \cite{Sype2004} to sense the input current during switch ON time. D-A control results in increased THD during Discontinuous Conduction Mode (DCM) operation. Trailing-Edge Modulated One Cycle Control controls the peak input current indirectly, by directly controlling the OFF-time of the switches. The input current's THD is higher than Average-Analog control \cite{Lu2005}. Digital-Peak Current samples the peak of the inductor's current, resulting in high THD when the input current's ripple is high.

The Leading-Edge Modulated One Cycle Control technique, differently from the D-PC, controls the ON-time of the switches. For light load, results present high total harmonic distortion and stability problems with low voltage gain. LEM-OCC Stability and Distortion and its simplified version are both able to control the ON-time of the switches with lower THD than the previous presented techniques and better stability results \cite{Fischer2020}. The THD for the LEM-OCC-SD is theoretically zero, although the generation of fictitious current in practice limits its THD, resulting in low but not zero values. LEM-OCC-SDS allows the use of analog circuits to implement the technique. 

A technique that presents stronger theoretical robustness than the previous commented techniques is switched control, since it guarantees Lyapunov-based stability for all designed operation range. From its switched nature, faster and more precise responses are expected than other control techniques, since the technique decides at which mode to operate based on the instantaneous value of the states, not on its mean value. The result of the controller is an operation mode, not a duty cycle value, which naturally avoids saturation. Also, as the system is modeled as a switched (affine) system, the differences between simulated results and real-world results are reduced. The switched control technique for affine systems is presented in the next section.

\section{Switched control for affine systems}
\label{sec:switched}
\begin{figure}[!t]
    \centering
    \begin{tikzpicture}[auto, node distance=1.5cm,>=latex']
        \node [input, name=input] {};
        \node [sum, right of=input] (sum) {};
        \node [block, right of=sum] (controller) {Controller};
        \node [block, right of=controller, node distance=3cm] (system) {System};
        \draw [->] (controller) -- node[name=u] {$\sigma$} (system);
        \node [output, right of=system] (output) {};
        \node [block, below of=u] (measurements) {Measurements};
    
        \draw [draw,->] (input) -- node[pos=0.2] {$x_{eq} \ \ -$} (sum);
        \draw [->] (sum) -- node {$e$} (controller);
        \draw [->] (system) -- node [name=y] {$x$}(output);
        \draw [->] (y) |- (measurements);
        \draw [->] (measurements) -| node[pos=0.9] {$+$} 
            node [near end] {} (sum);
    \end{tikzpicture}
    \caption{Control block diagram for the switched control technique.}
    \label{fig:controlblock}    
\end{figure}
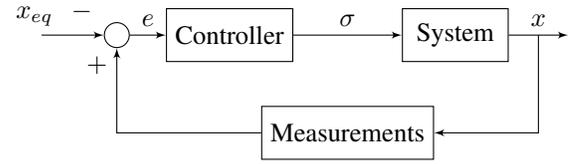

Considering the affine system with dynamics defined by \eqref{eq:affine}, a switching rule can be designed with the goal of driving the system state to a given constant equilibrium point $x_{eq}$. Given the tracking error dynamics
\begin{subequations}
    \begin{equation}
            \dot{e}(t) = A_i e(t) + k_i,
    \end{equation}
    \begin{equation}
            k_i = b_i + A_i x_{eq},        
    \end{equation}
    \begin{equation}
            e(t) \coloneqq x(t) - x_{eq},
    \end{equation}
\end{subequations}
a switching rule can be defined as \cite{Trofino2011}
\begin{subequations}
    \begin{equation}
        \sigma(e(t)) = \arg \max_{i \in \mathcal{M}} \{ v_i (e(t)) \},
    \end{equation}
    \begin{equation}
        v_i (e(t)) = e(t)' P_i e(t) + 2 e(t)' S_i,    
    \end{equation}
    \label{eq:switching_rule}
\end{subequations}
where $P_i = P_i' \in \mathbb{R}^{n \times n}$ and $S_i \in \mathbb{R}^n$ are the variables of decision, which need to be determined. Fig.~\ref{fig:controlblock} shows the block diagram for the switched control technique according to the aforementioned expressions. Sliding mode dynamics may occur in any switching surface and can be represented by
\begin{equation}
    \dot{e}(t) = \Sigma_{i=1}^{m} \theta_i (e(t)) (A_i e(t) + k_i), \ \theta(e(t)) \in \Theta,
    \label{eq:deriv_err}
\end{equation}
where $\Theta$ is the unitary simplex and $\theta_i (e(t)) = 0$ if $i \notin \sigma(e(t))$. 

From the main results obtained in \cite{Trofino2011}, it is supposed that $\exists P,S,L$ solving the Linear Matrix Inequality (LMI) problem
\begin{equation}
    \begin{cases}
        \bar{P} = \sum_{i=1}^{m} \bar{\theta}_i P_i > 0; \\
        \bar{S} = \sum_{i=1}^{m} \bar{\theta}_i S_i = 0; \\
        Q_a ' (\Psi + \Phi + L C_b(\theta) + C_b(\theta)' L') Q_a < 0,
    \end{cases}
    \label{eq:lmis}
\end{equation}
where
\begin{subequations}
    \begin{equation}
        \Psi = 
        \begin{bmatrix}
            A'P + P'A-\alpha'\bar{P}I_a-I_a'\bar{P}\alpha
            & \bullet \\
            K'P + S'A & K'S + S'K
        \end{bmatrix},
    \end{equation}
    \begin{equation}
        \Phi = 
        \begin{bmatrix}
            \alpha'(P - \bar{P} I_a) + (P' - I'_a \bar{P}) \alpha & \bullet \\ \\
            2 S' \alpha & 0_{m \times m}
        \end{bmatrix},
    \end{equation}
\end{subequations}    
with
\begin{subequations}
\label{eq:definitions}
    \begin{equation}
        A = [A_1 \ ...\  A_m]; \ \ \  K = [k_1 \ ... \ k_m],
    \end{equation}
    \begin{equation}
         P = [P_1 \ ... \ P_m]; \ \ \ S = [S_1 \ ... \ S_m],
    \end{equation}
    \begin{equation}
        \alpha = [\alpha_1 I_n \ ... \ \alpha_m I_n]; \ \ \ I_a = \mathbf{1}_m \otimes I_n,
    \end{equation}
    \begin{equation}
        \mathbf{1}_m = [1 \ ... \ 1] \in \mathbb{R}^m; \ \ \  C_a = [0_{1 \times m n} \ \ \mathbf{1}_m],
    \end{equation}
    \begin{equation}
        C_b(\theta) = 
        \begin{bmatrix}
            \aleph_{\theta} \otimes I_n & 0_{rn \times m} \\
            0_{r \times m} & \aleph_{\theta} - \aleph_{\bar{\theta}},
        \end{bmatrix},
    \end{equation}
\end{subequations}
in which $\bar{\theta}$ is a constant scalar such that
\begin{subequations}
    \begin{equation}
        \sum_{i=1}^{m} \bar{\theta}_i k_i = 0\ \ \ \ \sum_{i=1}^{m} \bar{\theta} = 1,
    \end{equation}
    \begin{equation}
        \bar{\theta} \geq 0, \ \ \ k_i = b_i + A_i x_{eq},
    \end{equation}
\end{subequations}
$\bullet$ represents block matrix terms that can be deduced from symmetry, $\aleph_{\theta}$ is the linear annihilator of $\theta$, $Q_a$ is a given basis for the null space of $C_a$, $L$ is a matrix to be determined with the dimensions of $Cb(\theta)'$. The closed-loop dynamics \eqref{eq:deriv_err} is globally assimptotically stable with the switching rule \eqref{eq:switching_rule} and
\begin{equation}
    V(e(t)) = \max_{i \in \mathcal{M}} \{ v_i (e(t)) \}
\end{equation}
is a Lyapunov function for the system, which guarantees Lyapunov-based stability.



\subsection{Uncertain systems}

The switched control technique can be extended to uncertain parameters of the system, as proposed in \cite{Trofino2009}, by considering a polytope with vertices $\{ \rho_1, \ ... \ , \rho_{2^d} \}$ for each uncertainty $\delta$. Assuming the desired $x_{eq}$ does not depend on $\delta$, the closed-loop dynamics \eqref{eq:deriv_err} becomes
\begin{equation}
    \dot{e}(t) = \Sigma_{i=1}^{m} \theta_i (e(t)) (A_i(\delta) e(t) + k_i(\delta)).
    \label{eq:deriv_err_unc}
\end{equation}

The standard convexity properties of the LMIs are held assuming $A_i(\delta)$ and $k_i(\delta)$ are affine functions of $\delta$, nevertheless the switched system \eqref{eq:deriv_err_unc}, now uncertain, is globally asymptotically stable for all uncertainties in the polytope if the LMIs are feasible $\forall \delta \in \{ \rho_1, \ ... \ , \rho_{2^d} \}$ and $i \in \mathcal{M}$ \cite{Trofino2009}.

\section{Extensions for power factor correction}
\label{sec:pfc}

As the operation of the TPBR can be divided in $v_{in} > 0$ (Modes 1 and 2) and $v_{in} < 0$ (Modes 3 and 4), the control problem can be approached separately for each sign of the input voltage. To avoid the need of obtaining $\bar{\theta}$, which is usually not feasible for this type of converter, all $P_i$ is considered equal within each case, $P_i \coloneqq P_0 \in \mathbb{R}^{n \times n}, \ i \in \mathcal{M}$. Also, in order to avoid the inclusion of $\sum_{i=1}^{m} \bar{S} = 0$ in the LMIs, which would also require to obtain $\bar{\theta}$, $v_i(e(t))$ may be changed from \eqref{eq:switching_rule} to
\begin{equation}
    v_i(e(t)) = e(t)' P_0 e(t) + 2 e(t)' (S_i - \bar{S}),
    \label{eq:vi_partial_theta}
\end{equation}
which makes $\bar{S}$ equal for all modes of each case. With the considerations stated above, knowing that only the terms that vary between modes are relevant to the switching, the switching law becomes
\begin{equation}
\label{eq:sw_implementation}
   \sigma(e(t)) = \arg \max_{i \in \mathcal{M}} \{ v_i(e(t)) \} = \arg \max_{i \in \mathcal{M}} \{ e(t)'S_i \}.
\end{equation}
Nevertheless, the LMI problem \eqref{eq:lmis} can be solved, for this case, as
\begin{equation}
    \begin{cases}
        P_0 > 0; \\
        Q_a ' (\Psi + \Phi + L C_b(\theta) + C_b(\theta)' L') Q_a < 0,
    \end{cases}
    \label{eq:lmis_pfc}
\end{equation}
where the following terms differ from \eqref{eq:definitions}:
\begin{equation}
    C_b(\theta) = 
    \begin{bmatrix}
        \aleph_{\theta} \otimes I_n & 0_{rn \times m}
    \end{bmatrix},
\end{equation}
\begin{equation}
    \Psi = 
    \begin{bmatrix}
        A'P + P'A
        & P' K + A' S \\
        K'P + S'A & K'S + S'K
    \end{bmatrix},
\end{equation}
\begin{equation}
    \Phi = 
    \begin{bmatrix}
        0_{rn^2 \times rn^2} & 2 \alpha' S \\
        2 S' \alpha & 0_{m \times m}
    \end{bmatrix}.
\end{equation}

With predefined values of maximum and minimum power for nominal operation, the load resistance can be taken into account as an uncertainty $\delta$ with the vertices of the polytope defined with $R_{L}^{min}$ and $R_{L}^{max}$. For power factor correction, the inductor's current $i_{L_B}$ must be controlled to have the same waveform of the input voltage $v_{in}$. That includes a new uncertainty to the LMIs allowing robustness for the whole range of $i_{L_B}$, defining the vertices of a polytope as: when $v_{in} > 0$, vertices are
\begin{equation}
    \label{eq:ilb_vertice1}
    i_{L_B} = 0 \quad \text{and} \quad i_{L_B} = \sqrt{2} \frac{P_o^{max}}{V_{in}^{rms}},
\end{equation}    
where $P_o^{max}$ is the maximum output power of the converter and $V_{in}^{rms}$ is the Root Mean Square (RMS) value of the minimum input voltage; and when $v_{in} < 0$, vertices are
\begin{equation}
\label{eq:ilb_vertice2}
    i_{L_B} = 0 \quad \text{and} \quad i_{L_B} = -\sqrt{2} \frac{P_o^{max}}{V_{in}^{rms}}.
\end{equation} 


\section{Simulation results}

\begin{table}[!t]
\renewcommand{\arraystretch}{1.3}
\caption{Parameters of the simulated TPBR.}
\label{tab:tpbr_parameters}
\centering
\begin{tabular}{c c c}
    \hline
    \textbf{Parameter} & \textbf{Description} & \textbf{Value} \\
    \hline
    $V_{in}^{rms,max}$ & Maximum RMS input voltage & 250 V \\
    $V_{in}^{rms}$ & Nominal RMS input voltage & 120 V \\
    $f_r$ & Line frequency & 60 Hz \\
    $L_B$ & Boost inductance & 2.4 mH \\
    $R_B$ & Boost inductor resistance & 0.42 $\Omega$ \\
    $P_o^{min}$ & Minimum output power & 25 W \\
    $P_o^{max}$ & Maximum output power & 300 W \\
    $V_o$ & Output voltage & 380 V \\
    $C_o$ & Output capacitance & 270 $\mu$F \\
    $f_s$ & Switching frequency considered for design & 64.8 kHz \\
    \hline
\end{tabular}
\end{table}

\label{sec:results}
The TPBR was simulated in Simulink\textregistered \ with nonideal components, with parameters shown in Table~\ref{tab:tpbr_parameters}, which were designed for the use of the converter for an inverter refrigerator \cite{Fischer2020}. The obtained solution $S_i$ for the LMI problem \eqref{eq:lmis_pfc} considering such specification, an $\alpha_i = 10000, \ \forall i \in \mathcal{M}$, is
\begin{subequations}
    \begin{equation}
        S_1 = 
        \begin{bmatrix}
            -0.2951 \\
            -0.0060
        \end{bmatrix}\times 10^{-5}  ,
    \quad
        S_2 = 
        \begin{bmatrix}
            0.5417 \\
            0.0000
        \end{bmatrix}\times 10^{-8} ,
    \end{equation}  
    \begin{equation}
        S_3 =
        \begin{bmatrix}
            0.3672 \\
            -0.0075
        \end{bmatrix} \times 10^{-5} ,
    \quad
        S_4 = 
        \begin{bmatrix}
            -0.5841 \\
            0.0000
        \end{bmatrix}\times 10^{-8},
    \end{equation}  
\end{subequations}
which was implemented in simulation according to \eqref{eq:sw_implementation}. Harmonic content was added to the input voltage altogether with white noise, in order to almost reach the THD limit of IEEE 519 \cite{IEEE519} for grid voltage (8 \%). This way, the controller is tested at the worst case scenario in terms of input voltage.

As the current reference signal is, in practice, generated with a Digital Signal Processor (DSP), two main options, in this case, should be considered for the reference signal: i) voltage's waveform - by sensing the input voltage and using its waveform as reference signal for the inductor's current, the power factor of the converter would be at its maximum. The downside is that the harmonic content of the input voltage would be inherited by the controlled input current, increasing its THD; ii) pure sinusoidal waveform - generating a pure sinusoidal signal for the current reference eliminates the need of sensing the whole waveform of the input voltage, only needing to measure its polarity. The obtained THD will be lower for this case. As input voltage and input current waveforms may differ, nonactive power will be present in the system. Therefore, the lower the distortion in the input voltage, the higher will be the obtained power factor.

\begin{figure*}
\begin{minipage}{0.33\linewidth}
    \centering
    \includegraphics[width=2.45in]{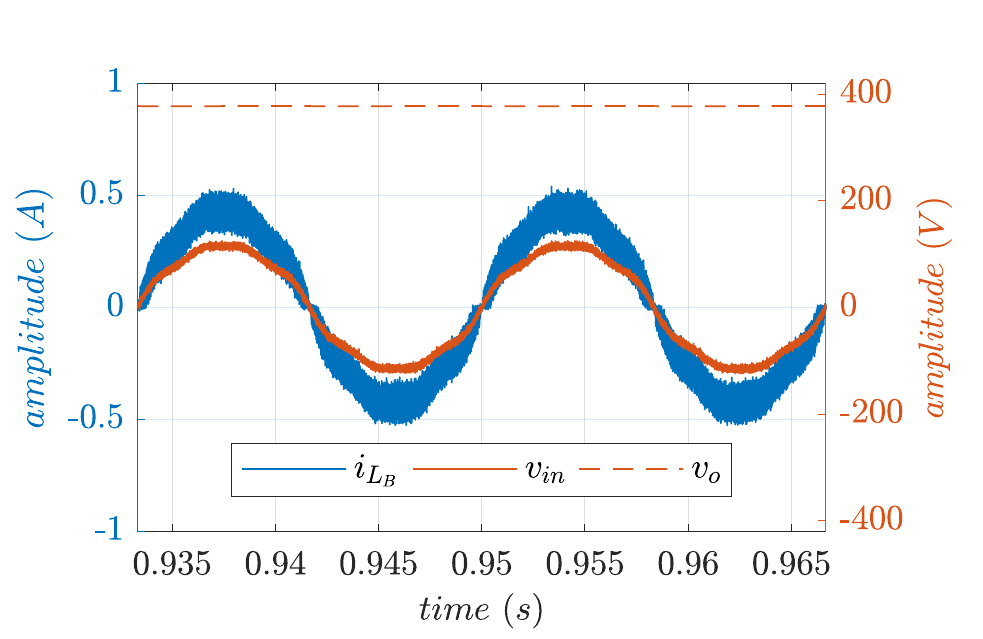}
    \subcaption{}
    \label{fig:85_25}
\end{minipage}
\begin{minipage}{0.33\linewidth}
    \centering
    \includegraphics[width=2.45in]{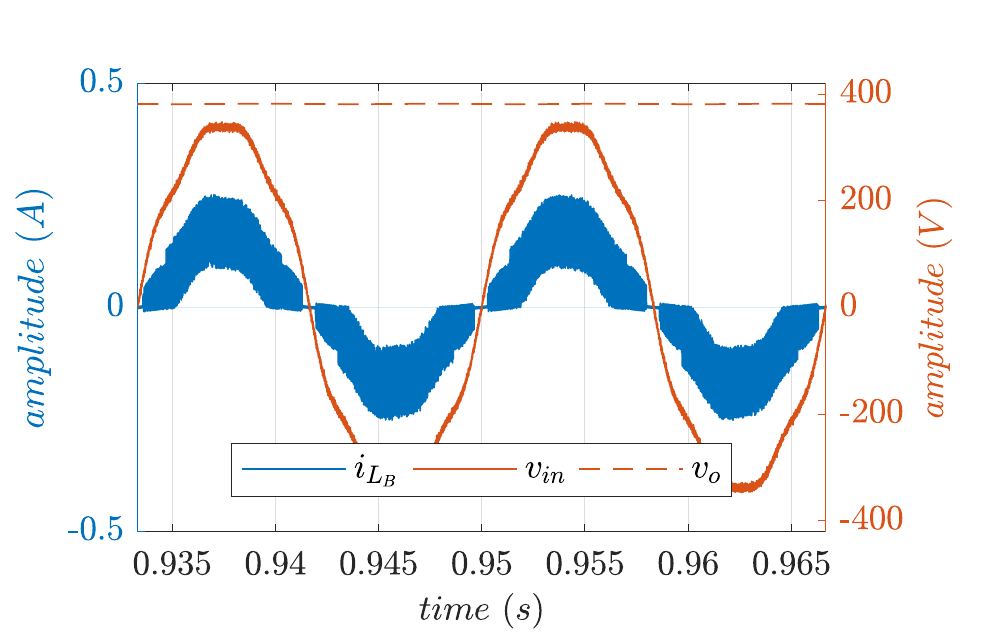}
    \subcaption{}
    \label{fig:250_25}
\end{minipage}
\begin{minipage}{0.33\linewidth}
    \centering
    \includegraphics[width=2.45in]{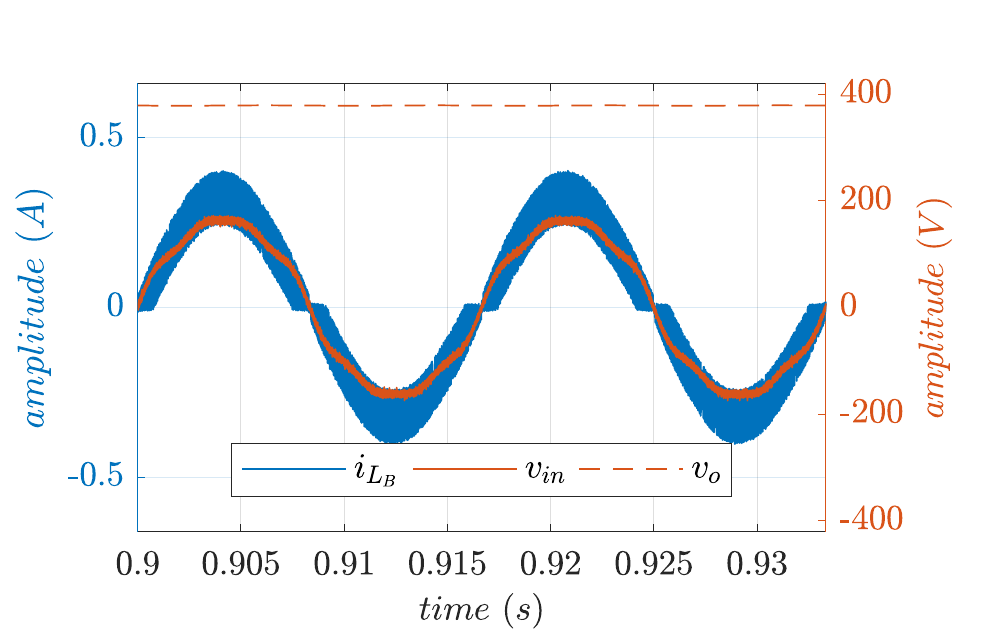}
    \subcaption{}
    \label{fig:120_25}
\end{minipage}
\caption{$i_{L_B}$, $v_{in}$ and $v_o$ for $P_o = 25$ W with: (a) $V_{in}^{rms} = 85$ V and input voltage waveform as current reference; (b) $V_{in}^{rms} = 250$ V and input voltage waveform as current reference; and (c) $V_{in}^{rms} = 120$ V and a pure sinusoidal current reference.}
\end{figure*}

\begin{figure*}
\begin{minipage}{0.33\linewidth}
    \centering
    \includegraphics[width=2.45in]{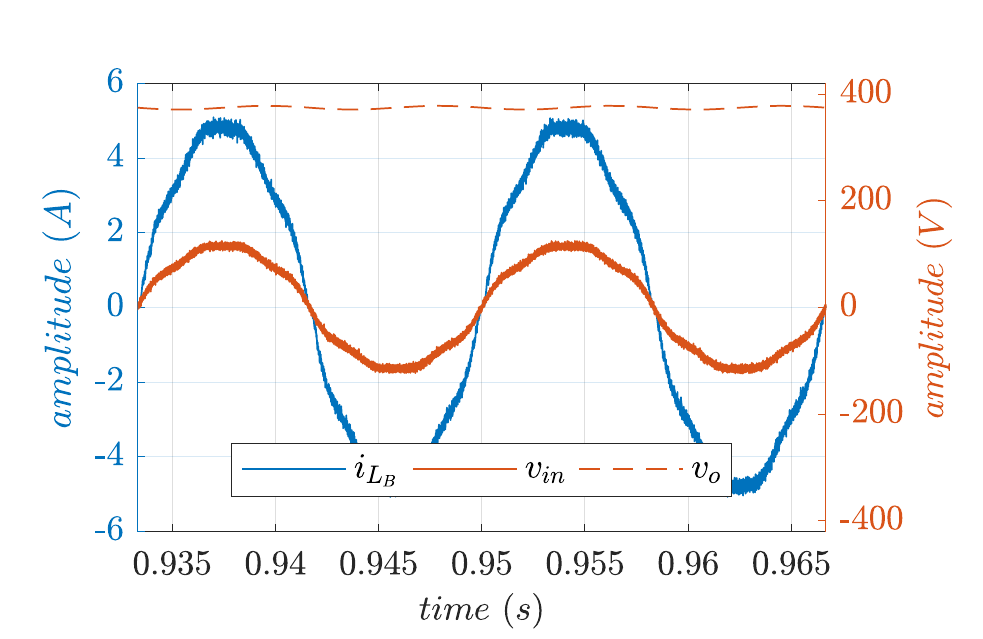}
    \subcaption{}
    \label{fig:85_300}
\end{minipage}
\begin{minipage}{0.33\linewidth}
    \centering
    \includegraphics[width=2.45in]{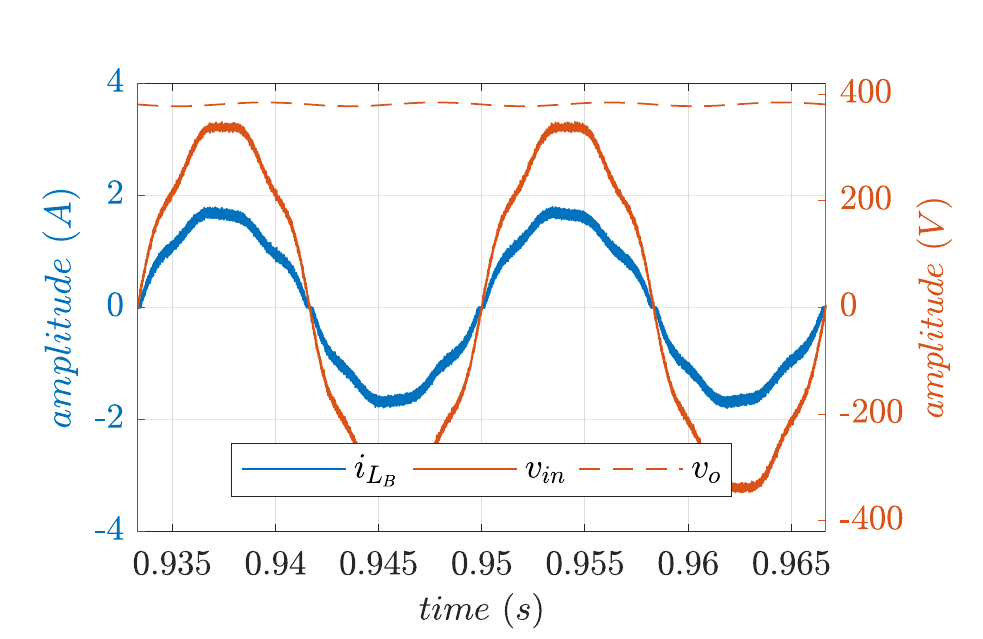}
    \subcaption{}
    \label{fig:250_300}
\end{minipage}
\begin{minipage}{0.33\linewidth}
    \centering
    \includegraphics[width=2.45in]{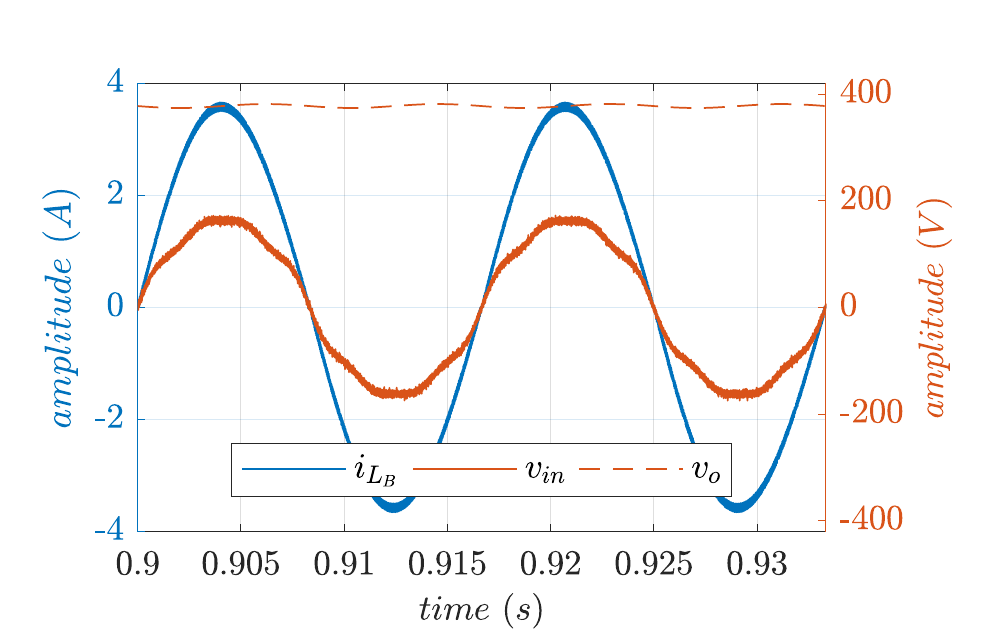}
    \subcaption{}
    \label{fig:120_300}
\end{minipage}
\caption{$i_{L_B}$, $v_{in}$ and $v_o$ for $P_o = 300$ W with: (a) $V_{in}^{rms} = 85$ V and input voltage waveform as current reference; (b) $V_{in}^{rms} = 250$ V and input voltage waveform as current reference; and (c) $V_{in}^{rms} = 120$ V and a pure sinusoidal current reference.}
\end{figure*}

Using the input voltage waveform as reference for the controller, regarding inductor's current, the vertices of the polytope were tested. Fig.~\ref{fig:85_25}~and~\ref{fig:85_300} shows the inductor's current $i_{L_B}$, input voltage $v_{in}$ and output voltage $v_o$ waveforms for $V_{in}^{rms} = 85$ V and output power $P_o$ of 25 and 300 W, respectively. Whilst, Fig.~\ref{fig:250_25}~and~\ref{fig:250_300} shows $i_{L_B}$, $v_{in}$ and $v_o$ for 250 V of input (rms) and $P_o = 25$ and $300$ W, respectively. The lower the output current, the greater will be the harmonic distortion in the input current waveform, which is a common behavior in PFC applications. Higher input voltage also results in higher THD in $i_{L_B}$. Note that when the input current is close to the zero value, which can be more clearly seen in Fig.~\ref{fig:85_25}~and~\ref{fig:250_25}, the ripple at the current's waveform is enough to touch the zero value, making the converter operate for a brief period of time in DCM. This phenomenon does not affect the switched control, since its range of operation regarding input current, see \eqref{eq:ilb_vertice1} and \eqref{eq:ilb_vertice2}, already includes the zero value. 

\begin{figure}[!t]
\centering
\includegraphics[width=3.3in]{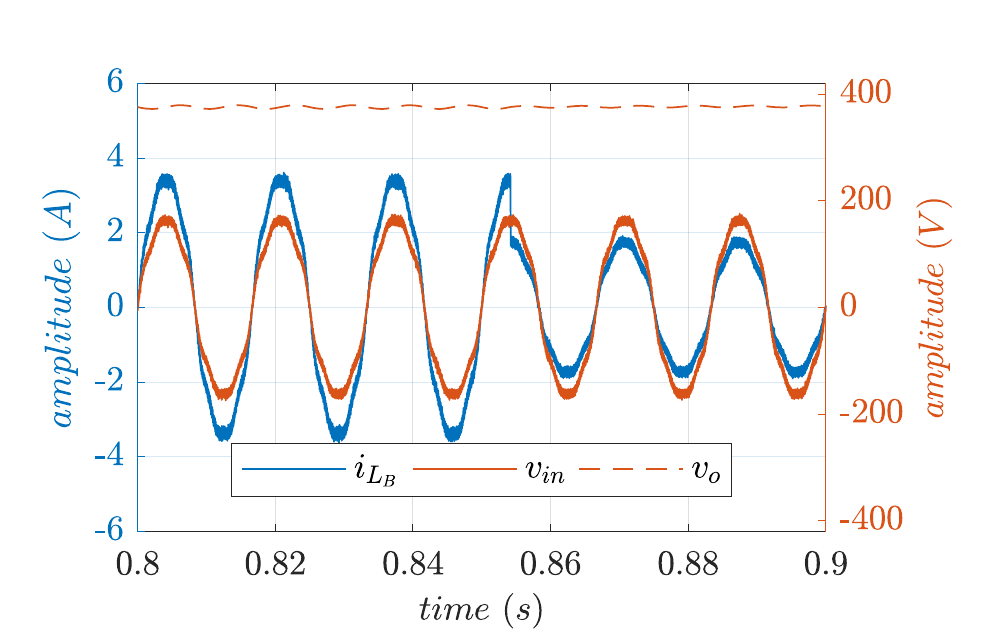}
\caption{Input voltage, inductor's current, and output voltage with a load step of half load at 0.854 seconds with the input voltage waveform as current reference.}
\label{fig:step}
\end{figure}

Fig.~\ref{fig:step} shows the input voltage, inductor's current and output voltage for $P_o^{max}$ with a load step to half of $P_o^{max}$ occurring at 0.854 seconds, the peak of the input current, with nominal input voltage $V_{in}^{rms}$ and distorted current reference waveform. No instability occurs during the load step, as it is included within the considered polytope in the LMIs, showing controller robustness. The controller's response is fast for both $i_{L_B}$ and $v_o$.

As aforementioned, the harmonic content of the input voltage waveform is inherited by the input current if its waveform is used as reference, increasing the current's THD. To demonstrate the reference following performance of the controller for another reference signal, a pure sinusoidal reference was applied to the controller. The input voltage is still distorted with harmonic content and corrupted by white noise, to the limits of IEEE 519. Two cases at the rated input voltage (120 V) are shown in Fig.~\ref{fig:120_25} and Fig.~\ref{fig:120_300}, for 25 and 300 W of output power respectively. As expected, the proposed technique can achieve a very precise reference following, lowering the THD and still maintaining high power factor.

\begin{figure}[!t]
\centering
\includegraphics[width=3.3in]{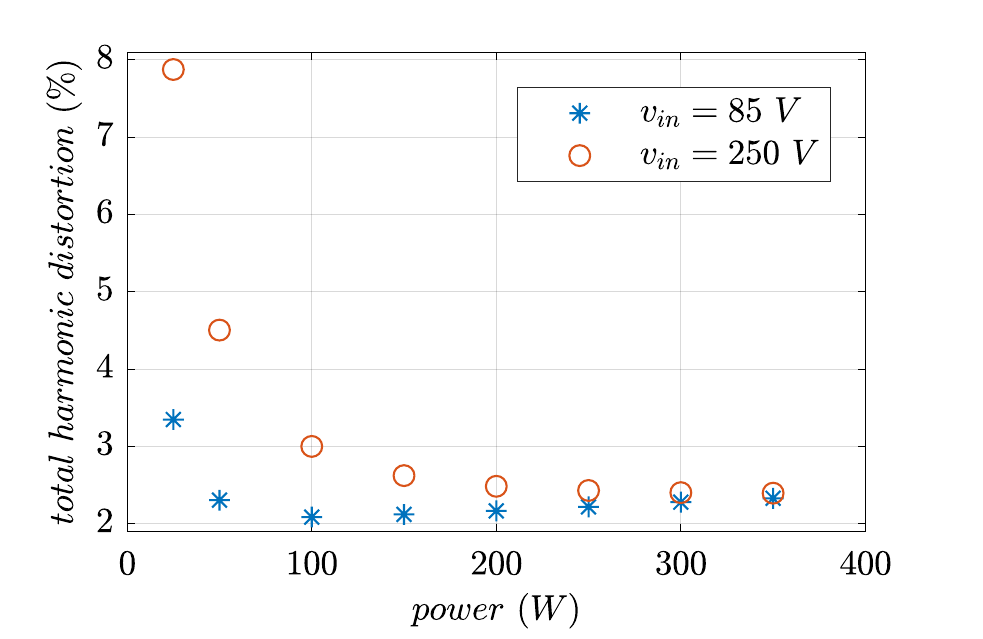}
\caption{Total harmonic distortion obtained for $v_{in} = 85$ V and $v_{in} = 250$ V with a pure sinusoidal current reference for different output power values.}
\label{fig:thd}
\end{figure}

Fig.~\ref{fig:thd} shows the values of total harmonic distortion of the inductor's current $i_{L_B}$ waveform for a pure sinusoidal reference. In comparison with values obtained in the literature \cite{Fischer2020}, it shows a significant improvement. The THD results confirm what is seen in Fig.~\ref{fig:85_25},~\ref{fig:85_300},~\ref{fig:250_25}~and~\ref{fig:250_300}: a higher THD for lower output current and for higher input voltage.

As the converter with specifications presented in Table~\ref{tab:tpbr_parameters} is designed for the application of household refrigerators, the normative energy quality requirements of IEC 61000-3-2 \cite{IEC61000} and JIS C 61000-3-2 \cite{JIS61000} are Class D. The presented solution for power factor correction with switched control is able to meet these requirements. The normative evaluate the THD at rated power, but switched control can obtain low THD at light load as well, which is a desired feature for household applications.

\section{Conclusion}
\label{sec:conclusions}
Switched control is a nonlinear control technique that, after designed, can be used in an offline fashion with low computational complexity, as it is applied in a digital signal processor device through linear matrices operations. This paper adapted switched control's theory to be applied for power factor correction in a Totem-Pole Bridgeless Rectifier.

The technique was applied through simulations in Simulink\textregistered \ with nonideal components and distorted input voltage, in a converter designed for household refrigeration with 300~W of nominal output power and input voltages from 85 to 250~V (rms). The results presented very low total harmonic
distortion for all tested cases and a fast and stable response for a load step from 100 to 50~\% of
load current. As the current reference waveform can be defined by the user through a DSP in practical applications, two current references were tested: the input voltage waveform; and a pure sinusoidal
waveform. Using the input voltage waveform as current reference, the converter achieves maximum power
factor, whilst for a pure sinusoidal current reference, it is possible to reduce the total harmonic
distortion.

As for a future works, it is suggested the application of this technique in a real prototype and its extension for partial measurement of the inductor's current. The latter should allow the user to measure a partial current at a low-side shunt resistor included between the output voltage's ground and the rectifier bridge formed by the diodes and transistors, reducing the cost of the sensors in the converter.

\bibliographystyle{IEEEtran}
\bibliography{conference_101719}


\end{document}